\documentclass[letterpaper, 10 pt, conference]{ieeeconf} 
\IEEEoverridecommandlockouts                             
\usepackage[english]{babel}
\usepackage{amsmath, amssymb, amsfonts, bm}
\usepackage{mathrsfs}
\usepackage{textcomp}
\usepackage[mathcal]{euscript}
\usepackage{enumerate}
\usepackage{mathbbol}
\usepackage{color,graphicx,epstopdf}
\usepackage{dsfont}
\usepackage[noadjust]{cite}
\usepackage{mathtools}
\usepackage{tikz-cd}
\usepackage{mathtools}
\usepackage{algorithm}
\usepackage{multirow}
\usepackage{algpseudocode}
\usepackage{url}
\usepackage{authblk}
\usepackage{verbatim}
\usepackage{stackengine}
\usepackage{subcaption}

\newtheorem{theorem}{Theorem}

\newtheorem{remark}{Remark}

\newtheorem{assumption}{Assumption}

\newtheorem{experiment}{Experiment}
\usepackage[left=1.91cm,right=1.91cm,top=1.91cm,bottom=1.91cm]{geometry}
\usepackage{float}

\renewcommand{\b}[1]{\mathbf{#1}}
\newcommand{\bs}[1]{{\boldsymbol #1}}

\title{\LARGE \bf
	 Using Gaussian Mixtures to Model Evolving \\ Multi-Modal Beliefs Across Social Media
}

\author{Yijun Chen$^{1}$, Farhad Farokhi$^{1}$, Yutong Bu$^{1, 3}$, Nicholas Kah Yean Low$^{2, 3}$, Jarra Horstman$^{2}$,\\ Julian Greentree$^{1}$, Robin Evans$^{1}$, Andrew Melatos$^{2, 3}$
	
	\thanks{$^{1}$Department of Electrical and Electronic Engineering, The University of Melbourne, Parkville, VIC 3010, Australia.}%
	\thanks{$^{2}$School of Physics, The University of Melbourne, Parkville, VIC 3010, Australia.}
    \thanks{$^{3}$Australian Research Council Centre of Excellence for Gravitational Wave Discovery, The University of Melbourne, Parkville, VIC 3010, Australia.}
    \thanks{Email: \{yijun.chen.1, farhad.farokhi, robinje, amelatos\}@unimelb.edu.au (Chen, Farokhi, Evans,  Melatos), \{y.bu2, nlow1, jhorstman\}@student.unimelb.edu.au (Bu, Low, Horstman), and greentreejj@gmail.com (Greentree).}
	\thanks{This work was supported by the Australian Research Council under grants CE170100004. The work of  Bu is supported by a Research Training Program Scholarship from Australian Commonwealth Government and The University of Melbourne. The work of Low is supported by a Research Training Program Scholarship from Australian Commonwealth Government and The University of Melbourne, and the Robert D Hill Scholarship.}
}

\begin{document}

\maketitle
\thispagestyle{empty}
\pagestyle{empty}

\begin{abstract}
We use Gaussian mixtures to model formation and evolution of multi-modal beliefs and opinion uncertainty across social networks. In this model, opinions evolve by Bayesian belief update when incorporating exogenous factors (signals from outside sources, e.g., news articles) and by non-Bayesian mixing dynamics when incorporating endogenous factors (interactions across social media).
The modeling enables capturing the richness of behavior observed in multi-modal opinion dynamics while maintaining interpretability and simplicity of scalar models. We present preliminary results on opinion formation and uncertainty to investigate the effect of stubborn individuals (as social influencers). This leads to a notion of centrality based on the ease with which an individual can disrupt the flow  of information across the social network.
\end{abstract}

\section{Introduction}\label{sec:introduction}    
Opinion formation and information exchange on social media can have a direct impact on election outcomes~\cite{druckman2005impact}, social cohesion~\cite{gonzalez2023social}, and public health~\cite{viswanath2007mass}. Therefore, modeling opinion dynamics on complex social networks and predicting the spread of misinformation are valuable. Mathematical models of opinion dynamics, belief propagation, and social learning facilitate this inquiry by capturing opinion formation and evolution based on exogenous factors, such as news and political events, and endogenous factors, such as social interaction and peer pressure.

A category of opinion dynamics models follows the pioneering work of French~\cite{french1956formal}, DeGroot~\cite{degroot1974reaching}, and Friedkin-Johnsen~\cite{friedkin1997social} by modeling an individual's belief with a deterministic scalar variable.~That is, an individual can only hold a single belief with certainty and cannot hold multiple beliefs with varying levels of certainty (cf.~an undecided voter). These models have proven powerful in the analysis of media and disinformation~\cite{carletti2006make,pineda2015mass}, opinion polarization~\cite{amelkin2017polar,baumann2020modeling}, and social power structure~\cite{9992374,jia2015opinion}. However, a valid criticism of these opinion-dynamics models is that they do not account for uncertainty in belief and for holding multiple opinions. 

Another approach is to use multi-modal opinions that incorporate uncertainty using probability distributions~\cite{low2022discerning,anunrojwong2018naive, jadbabaie2012non, bu2023discerning}.~In this framework, opinions evolve by Bayesian belief update when incorporating exogenous factors and by Bayesian or non-Bayesian mixing dynamics when incorporating endogenous factors. The complex interplay between rich internal models for individuals and dissemination of information across network topology enables these models to capture a wide array of behaviors. However, these models stray from the simple form of DeGroot or Friedkin-Johnsen models by incorporating
high-dimensional probability distributions to model beliefs, which makes them less amenable to analytic analysis. The dimension of the model can depend on discretization resolution
rather than intrinsic complexity of opinion formation (e.g., despite possessing a high discretization resolution, it is observed that bi-modal opinions form rapidly in the presence of partisans~\cite{bu2023discerning}). Instead of direct discretization, one can restrict the set of probability density functions utilized to model beliefs and track meaningful parameters for that specific set. For instance, Gaussian probability distributions can lead to DeGroot-like updates and thus uni-modal belief propagation~\cite{golub2010naive}. 

This paper fills the gap of using Gaussian mixture to model multi-modal opinion dynamics and uncertainty in social network, where individuals hold multiple, often conflicting beliefs shaped by exogenous observation signals (e.g., news articles) and endogenous social interactions over network. This method enables us to capture the richness of behavior observed in multi-dimensional opinion dynamics~\cite{low2022discerning, bu2023discerning} while maintaining interpretability, simplicity, and intuitiveness of scalar models~\cite{degroot1974reaching,golub2010naive} that primarily stems from dealing with Gaussian beliefs. We then propose a novel framework integrating Bayesian belief updates for external observation integration and non-Bayesian mixing dynamics for peer influence over network, grounded in Gaussian mixtures to capture opinion multi-modality and uncertainty. Additionally, we also investigate the effect of partisan or stubborn agents. Our key results include: (1) conditions under which truth discovery emerges (``wisdom of crowds'') despite conflicting signals and social biases; (2) a centrality measure quantifying the influence of stubborn or partisan individuals in manipulating information flow.

The remainder of the paper is organized as follows. Section~\ref{sec:formulation} formulates the opinion dynamics model using Gaussian mixtures, unifying Bayesian updates for external signals and non-Bayesian rules for social interactions. Section~\ref{sec:mainresult} establishes theoretical conditions for truth discovery. Section~\ref{sec:stubborn} introduces a stubborn agent as a social influencer, and proposes a centrality metric to measure its disruptive potential. Section~\ref{sec:simulation} carries out extensive numerical experiments. Finally, Section~\ref{sec:conc} concludes with future directions.

\subsection{Notations}
We denote random variables using boldface lowercase letters, both Roman and Greek (e.g., $\mathbf{x}$ and $\bs{\theta}$). The specific values these random variables can take are represented by normal lowercase letters, both Roman and Greek (e.g., $x$ and $\theta$).

A Gaussian random variable $\mathbf{x}$ with mean $\mu_{\mathbf{x}}$ and variance $\sigma_{\mathbf{x}}$ follows the probability density function:
\[
\mathcal{N}(x \mid \mu_{\mathbf{x}}, \sigma_{\mathbf{x}}) := \frac{1}{\sqrt{2\pi\sigma_{\mathbf{x}}}} \exp\left(-\frac{(x - \mu_{\mathbf{x}})^2}{2\sigma_{\mathbf{x}}}\right).
\]
To simplify notation in subsequent equations, we use $\sigma_{\mathbf{x}}$ to denote the variance, rather than the traditional $\sigma_{\mathbf{x}}^2$. We write $\mathbf{x} \sim \mathcal{N}(\mu_{\mathbf{x}}, \sigma_{\mathbf{x}})$ to indicate that random variable $\b{x}$ follows a Gaussian distribution with the specified parameters.

While Gaussian distributions are powerful for modeling uni-modal data, many real-world phenomena exhibit more complex, multi-modal patterns. 
In statistics literature, the Gaussian framework has been extended to Gaussian-mixture distributions, which combines multiple Gaussian components into a single probabilistic model to capture such complexity. 

\begin{figure}[bt]
    \centering
    \includegraphics[width=0.95\linewidth]{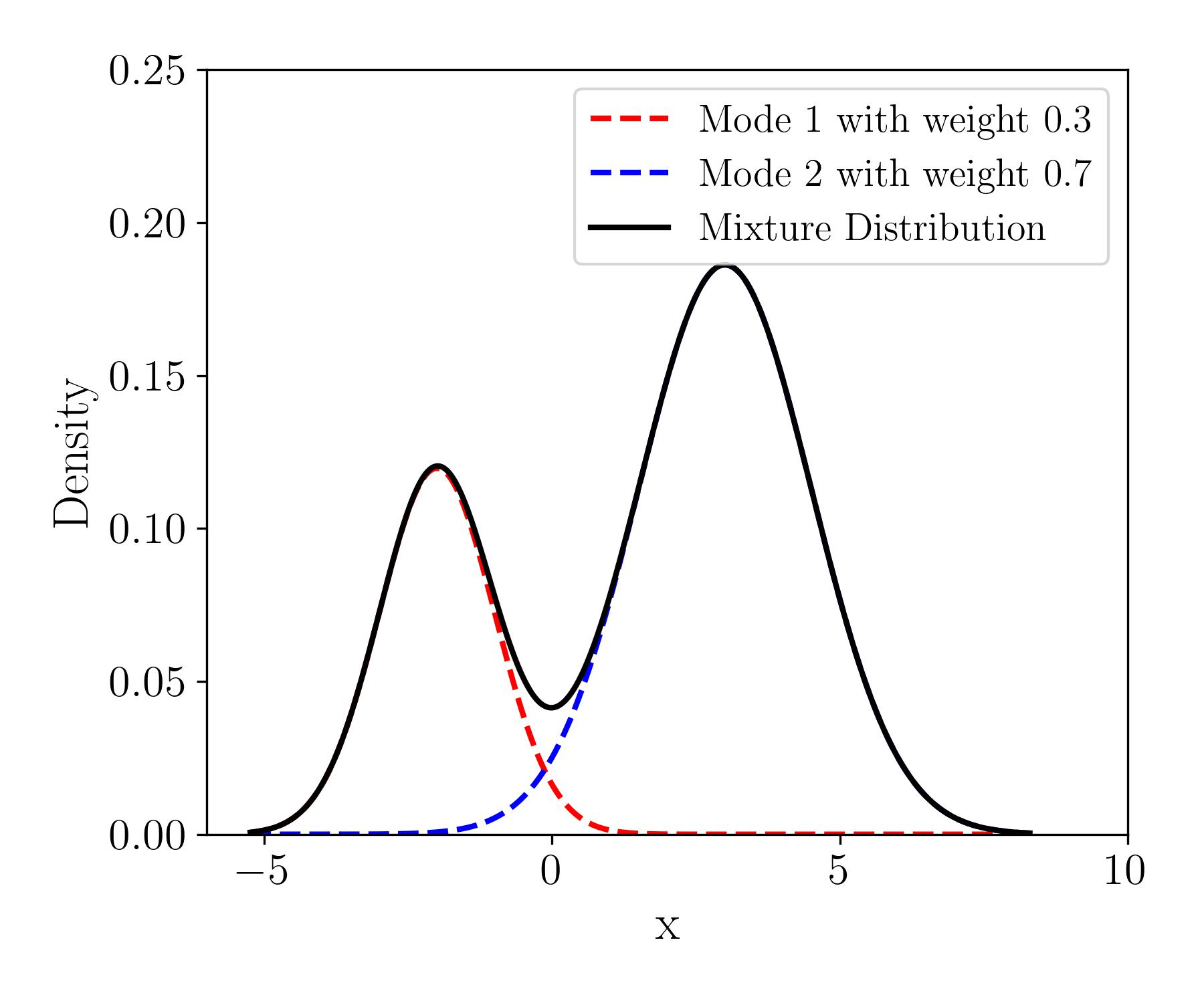}
    \caption{The probability density function of a Gaussian mixture model with two modes. The first Gaussian mode has a mean of \( \mu^{(1)}_{\mathbf{x}} = -2 \), a standard deviation of \( \sigma^{(1)}_{\mathbf{x}} = 1 \), and a weight of \( \alpha^{(1)}_{\mathbf{x}} = 0.3 \) (red dashed line). The second mode has a mean of \( \mu^{(2)}_{\mathbf{x}} = 3 \), a standard deviation of \( \sigma^{(2)}_{\mathbf{x}} = 1.5 \), and a weight of \( \alpha^{(2)}_{\mathbf{x}} = 0.7 \) (blue dashed line). The overall Gaussian mixture distribution is represented by the black solid line.
}
    \label{fig:GMM}
\end{figure}

A Gaussian-mixture random variable $\mathbf{x}$ with $M$ modes is characterized by means $\{\mu_{\b{x}}^{(i)}\}_{i=1}^M$, variances $\{\sigma_{\b{x}}^{(i)}\}_{i=1}^M$, and weights $\{\alpha_{\b{x}}^{(i)}\}_{i=1}^M$, where the weights satisfy $\sum_{i=1}^M \alpha_{\b{x}}^{(i)}=1$. Its probability density function is given by
\begin{align*}
    \mathcal{M}(x\mid \{\mu_{\b{x}}^{(i)}\}_{i=1}^M,&\{\sigma_{\b{x}}^{(i)}\}_{i=1}^M, \{\alpha_{\b{x}}^{(i)}\}_{i=1}^M)\\
    :=&
    \sum_{i=1}^M\alpha_{\b{x}}^{(i)} \frac{1}{\sqrt{2\pi\sigma_{\b{x}}^{(i)}}}
    \exp\left(-\frac{(x-\mu_{\b{x}}^{(i)})^2}{2\sigma_{\b{x}}^{(i)}}\right)\\
    =&
    \sum_{i=1}^M\alpha_{\b{x}}^{(i)} 
    \mathcal{N}(x|\mu_{\b{x}}^{(i)},\sigma_{\b{x}}^{(i)}).
\end{align*}
Similarly, $\b{x}\sim\mathcal{M}(\{\mu_{\b{x}}^{(i)}\}_{i=1}^M,\{\sigma_{\b{x}}^{(i)}\}_{i=1}^M,\{\alpha_{\b{x}}^{(i)}\}_{i=1}^M)$ indicates that random variable $\b{x}$ follows a Gaussian-mixture distribution with the specified parameters. An example of a  two-modal Gaussian mixture distribution is depicted in Fig.~\ref{fig:GMM}. 

\section{Model Description}\label{sec:formulation}
Consider a social network consisting of $N$ agents, which is represented by a weighted directed graph $\mathcal{G} := (\mathcal{V}, \mathcal{E}, W)$. Here, $\mathcal{V} = \{1, \dots, N\}$ represents the set of nodes corresponding to the agents, $\mathcal{E} \subseteq \mathcal{V} \times \mathcal{V}$ denotes the set of edges representing the connections between agents, and $W \in \mathbb{R}^{N \times N}$ is the weight matrix encoding the intensity of these connections. An edge $(i, j) \in \mathcal{E}$ indicates that agent~$i$ can influence agent $j$, where the influence intensity is quantified by the weight $w_{ij}$. Note that $w_{ij} = 0$ if and only if $(i, j) \notin \mathcal{E}$.

\subsection{Media Bias, Agent Prior Beliefs, and Agent Observations}
In the considered social network, a news media organization is characterized by an inherent scalar bias parameter $\theta \in \mathbb{R}$ representing its ideological leaning. For the sake of interpretation, $\theta = -\infty$ corresponds to extreme left, $\theta = +\infty$ corresponds to extreme right, and $\theta = 0$ corresponds to centrism. 

Each agent holds the prior belief that the news media organization's  bias  is best modeled by  Gaussian mixtures to capture multi modality of belief. The mode index set is denoted by $\mathsf{M} = \{1,2,\dots, M\}$. For instance, agents may perceive the news media organization in two conflicting ways. They may believe it is right-leaning with a probability of $30\%$ and left-leaning with a probability of $70\%$.
\begin{assumption}[Gaussian-mixture beliefs]\label{assum:belief}
    At time $k=0$, each agent $j\in \mathcal{V}$ holds the prior belief that $\bs{\theta}\sim\mathcal{M}(\{\mu^{(i)}_j[0]\}_{i=1}^M,\{\sigma^{(i)}_j[0]\}_{i=1}^M,\{\alpha^{(i)}_j[0]\}_{i=1}^M)$.
\end{assumption}
\medskip

The news media organization releases articles for public consumption over time, and the agents' observations of the news media organization’s ground truth bias at time $k \in \mathbb{N}$ homogeneously follow a Gaussian distribution. 

\begin{assumption}[Gaussian Observation]
At time $k\in\mathbb{N}$, an observation $\b{y}[k]$ is measured such that, conditioned on the ground truth bias $\bs{\theta}$, $\b{y}[k]\sim \mathcal{N}(\bs{\theta},\sigma_{\b{y}})$. 
\end{assumption}

This assumption, in essence, means that the observation $\b{y}$ is equal to true media bias plus a zero-mean Gaussian random variable with variance $\sigma_{\b{y}}$. It can also be extended to Gaussian-mixture likelihoods by following the results of~\cite{flam2012mmse}. However, we focus on Gaussian likelihoods in this paper for the sake of exposition and simplicity.

\begin{remark}
The potential bias in the observations is not a significant concern for two reasons. First, all the following analysis can be easily modified to account for biased observations. Second, if the observations are always biased by a fixed amount, this bias can be absorbed into the media bias itself, as it essentially acts as an additional component of the media bias.
\end{remark}

\subsection{Belief Update Rules}
Agents would like to estimate the ground truth bias $\bs{\theta}$ of the news media organization based on two factors. The first is an exogenous factor—their observations of sample articles from the organization, which forms the Bayesian belief update. The second is an endogenous factor—their belief exchanges over the social network, which leads to the social belief update.

\subsubsection{Bayesian Update}
In what follows, we show that the Gaussian-mixture form of the beliefs is preserved at every time step after observations.

To be specific, suppose that, at each time step $k\in\mathbb{N}$, agent $j \in \mathcal{V}$ holds a Gaussian-mixture prior $\bs{\theta}\sim\mathcal{M}(\{\mu^{(i)}_j[k]\}_{i=1}^M,\{\sigma^{(i)}_j[k]\}_{i=1}^M,\{\alpha^{(i)}_j[k]\}_{i=1}^M)$. Meanwhile, the news media organizations release articles that provide Gaussian observations $\b{y}[k]\sim\mathcal{N}(\bs{\theta},\sigma_{\b{y}})$ to the agents. Naturally, each agent performs a Bayesian update to refine its belief, which can be derived directly from Bayes’ theorem (see \cite{4518754}). Since the priors are Gaussian-mixture and the observations are Gaussian, the posteriors remain Gaussian-mixture. The Gaussian-mixture parameters are updated simultaneously as follows
\begin{subequations}\label{eq:bayesian_update}
    \begin{align}
    \mu^{(i)+}_j[k]
    &=\mu^{(i)}_j[k]
    +\frac{\sigma^{(i)}_j[k]}{\sigma^{(i)}_j[k]+\sigma_{\b{y}}}(\b{y}[k]-\mu^{(i)}_j[k]), \label{eq:mu_posterior_update}\\
    \sigma^{(i)+}_j[k]
    &=\sigma^{(i)}_j[k]
    -\frac{{\sigma^{(i)}_j[k]}^2}{\sigma^{(i)}_j[k]+\sigma_{\b{y}}}, \label{eq:sigma_posterior_update}\\
    \alpha^{(i)+}_j[k]
    &=\frac{\alpha^{(i)}_j[k] \cdot \mathcal{N}(\b{y}[k]|\mu^{(i)}_j[k],\sigma^{(i)}_j[k])}{\sum_{i=1}^M \big(\alpha^{(i)}_j[k] \cdot \mathcal{N}(\b{y}[k]|\mu^{(i)}_j[k],\sigma^{(i)}_j[k])\big)},\label{eq:alpha_posterior_update}
\end{align}
\end{subequations}
for all $j \in \mathcal{V}$, $k \in \mathbb{N}$, and $i\in \mathsf{M}$.

\begin{remark}[Relationship to discrete distributions] 
If we fix (and do not update) the location $\{\mu^{(i)}_j[k]\}_{i=1}^M$ and width of the Gaussian modes $\{\sigma^{(i)}_j[k]\}_{i=1}^M$ in our Gaussian-mixture model, we can think of each Gaussian mode as a ``bin'' and the weights $\{\alpha^{(i)}_j[k]\}_{i=1}^M)$ as the probability associated with that bin. This is very similar to the discrete model used in~\cite{bu2023discerning,low2022discerning,low2022vacillating}.
Here, our modeling and Bayesian updates also change the location and width of the Gaussian modes. Therefore, the framework can be considered an extension of discrete models.
\end{remark} 

\subsubsection{Social Update}

After performing the Bayesian update at each time step $ k \in \mathbb{N} $, agents exchange information through the social network and adjust their beliefs socially. The social belief update process is inspired by opinion dynamics models \cite{bu2023discerning,golub2010naive,low2022discerning,low2022vacillating,friedkin1997social,degroot1974reaching} and is described by the 
following update equations:
\begin{subequations}\label{eq:social_update}
\begin{align}
    \mu^{(i)}_j[k+1] &= \mu^{(i)+}_j[k] + f_{\mu}(\mu^{(i)+}_j[k],\mu^{(i)+}_{-j}[k]),\label{eq:mu_social_update}\\
    \sigma^{(i)}_j[k+1] &= \sigma^{(i)+}_j[k] + f_{\sigma}(\sigma^{(i)+}_j[k],\sigma^{(i)+}_{-j}[k]),\label{eq:sigma_social_update}\\
    \alpha^{(i)}_j[k+1] &= \alpha^{(i)+}_j[k] \cdot f_{\alpha}(\alpha^{(i)+}_j[k],\alpha^{(i)+}_{-j}[k]),\label{eq:alpha_social_update}
\end{align}
\end{subequations}
for all $ j \in \mathcal{V} $, $ k \in \mathbb{N} $, and $ i \in \{1, \dots, M\} $. Here, $ (\cdot)_{-j} \in \mathbb{R}^{N-1} $ represents the corresponding elements of all agents except agent $ j $. The functions of social update policies $ f_{\mu}: \mathbb{R}^{N} \to \mathbb{R} $, $ f_{\sigma}: \mathbb{R}^{N} \to \mathbb{R} $, and $ f_{\alpha}: \mathbb{R}^{N} \to \mathbb{R} $ are continuous functions that capture the influence of the social network on the agents' belief evolution.  It is noted that the weights $\alpha^{(i)}_j[k+1], i \in {1,\cdots,M}$ of each agent $j \in \mathcal{V}$ at each time step $k \in \mathbb{N}$ needs to satisfy the constraint $\sum_{i = 1}^{M}\alpha^{(i)}_j[k+1] = 1.$   In what follows, we investigate a few basic social update policies $f_{\mu}, f_{\sigma}, f_{\alpha}$ to establish some preliminary results.


\section{Convergence Analysis}\label{sec:mainresult}

In this section, we present convergence results for belief update rules.

\subsection{The Variance Dynamics}
In this subsection, we look at the variance dynamics. We first present a result when there is no network bias, i.e., $f_{\sigma} = 0$, when variables take identical values, i.e., the problem is initialized symmetrically.
\begin{theorem}\label{thm:IAA}
Assume that if $x_1 = x_2 = \dots = x_N$, 
then $f_{\sigma}(x_1, x_2, \dots, x_N) = 0$. Also assume that $\sigma_j^{(i)}[0]=\sigma_{0}>0$ for all $i\in \mathsf{M}$ and $j\in \mathcal{V}$. If $\sigma_{\b{y}}>0$,
\begin{align*}
    \sigma^{\infty}:=\lim_{k\rightarrow \infty}\sigma_j^{(i)}[k]=0, \quad \forall i\in \mathsf{M}, \forall j\in \mathcal{V}.
\end{align*}

\end{theorem}
\medskip

\begin{proof}
    We first note that each agent's initial variance is the same, i.e., 
    $\sigma_{j}^{(i)}[0] = \sigma_{0}$ for all $i \in  \mathsf{M}$ and $j \in \mathcal{V}$. 
    By the property that $f_{\sigma}=0$ when the variables are identical 
    and the social update equations~\eqref{eq:sigma_social_update}, 
    we have
    \[
        \sigma_{j}^{(i)}[k+1] = \sigma_{j}^{(i)+}[k],
    \]
    for all $\forall j \in \mathcal{V}, k \in \mathbb{N}, \text{ and } i\in \mathsf{M}.$
    Next, due to the equal initial variances and the symmetry of the Bayesian update equations~\eqref{eq:sigma_posterior_update}, 
    it leads to equal posterior variances
    \[
        \sigma_j^{(i)+}[k] = \sigma[k], 
    \]
    for all $\forall j \in \mathcal{V}, k \in \mathbb{N}, \text{ and } i\in \mathsf{M}.$
    Combining the above equations, we have
    \[
        \sigma[k+1] = \sigma[k] \,\frac{\sigma_{\mathbf{y}}}{\sigma[k] + \sigma_{\mathbf{y}}},
    \]
    for all $k \in \mathbb{N}$. We compare $\sigma[k+1]$ with $\sigma[k]$:
    \[
        \sigma[k+1] - \sigma[k]
        = -\,\frac{\sigma[k]^{2}}{\sigma[k] + \sigma_{\mathbf{y}}},
    \]
    which is strictly negative for all $\sigma[k] > 0$. 
    Hence, the sequence $\{\sigma[k]\}_{k \in \mathbb{N}}$ is strictly decreasing 
    and is bounded below by $0$. Therefore, it must converge. 
    Taking limits on both sides of the update equation yields
    \[
        \sigma^{\infty} 
        \;=\; \sigma^{\infty} \,\frac{\sigma_{\mathbf{y}}}{\sigma^{\infty} + \sigma_{\mathbf{y}}}.
    \]
    This equality admits the unique solution $\sigma^{\infty} = 0$. 
    This completes the proof. 
\end{proof}

\medskip
\begin{remark}
    The symmetric initial condition $\sigma_j^{(i)}[0] = \sigma > 0$ is not especially restrictive. It simply implies that all agents begin with the same positive level of uncertainty about their beliefs—one might say they are ``equally confused'' initially. 
\end{remark}
\medskip




We next present a result when there is network bias, i.e., $f_{\sigma} = \nu$, when variables take identical values.

\begin{theorem}\label{thm:BAA}
Assume that if $x_1 = x_2 = \dots = x_N$, 
then $f_{\sigma}(x_1, x_2, \dots, x_N) = \nu$. Also assume that $\sigma_j^{(i)}[0]=\sigma_{0}>0$ for all $i\in \mathsf{M}$ and $j\in \mathcal{V}$. If $\sigma_{\b{y}}>0$,
\begin{align}\label{eq:sigma_infty}
    \sigma^{\infty}:=\lim_{k\rightarrow \infty}\sigma_j^{(i)}[k]=\frac{\nu+ \sqrt{\nu^2+4\nu\sigma_{\b{y}}}}{2}, 
\end{align}
for all $i\in \mathsf{M}$ and all $j\in \mathcal{V}$.
\end{theorem}
\medskip

\begin{proof}
    Again, due to the equal initial variances and the symmetry of the Bayesian update equations~\eqref{eq:sigma_posterior_update}, 
    it leads to equal posterior variances $\sigma_j^{(i)+}[k]=\sigma[k]$     for all $\forall j \in \mathcal{V}, k \in \mathbb{N}, \text{ and } i\in \mathsf{M}$. Further combining the social update equations~\eqref{eq:sigma_social_update}, we have
    \begin{align*}
        \sigma[k+1]=\sigma[k] \frac{\sigma_{\b{y}}}{\sigma[k]+\sigma_{\b{y}}}+\nu.
    \end{align*}
    Consider the recurrence recurrence relation $\sigma[k+1] = g(\sigma[k])$, where the function $g(x)$ is defined as
    \begin{align*}
        g(x):=x\frac{\sigma_{\b{y}}}{x+\sigma_{\b{y}}}+\nu,
    \end{align*}
    with $\sigma_{\b{y}} > 0 $ and $\nu >0$. We first observe that if $x\in[\nu,\infty)$, then $g(x)\in[\nu,\infty)$. Next, we want to show that $g$ is a contraction on $[\nu, \infty)$. Consider $x,y \in [\nu, \infty)$:
    \begin{align*}
        |g(x)-g(y)|
        &=\left|x \frac{\sigma_{\b{y}}}{x+\sigma_{\b{y}}}-y\frac{\sigma_{\b{y}}}{y+\sigma_{\b{y}}}\right|
        \\
        &=\left| \frac{\sigma_{\b{y}}^2}{(x+\sigma_{\b{y}})(y+\sigma_{\b{y}})}\right||x-y|\\
        &\leq \frac{\sigma_{\b{y}}^2}{(\nu+\sigma_{\b{y}})^2}|x-y|,
    \end{align*}
    where $\frac{\sigma_{\b{y}}^2}{(\nu+\sigma_{\b{y}})^2} < 1.$
    It indicates that $g$ is a contraction. Therefore, from the Banach fixed point theorem, $g$ must admit a unique fixed point $\sigma^{\infty} \in [\nu, \infty)$ and $\{\sigma[k]\}_{k\in\mathbb{N}\setminus\{1\}}$ converges to $\sigma^{\infty}$. 
    To find the fixed point, we solve $\sigma^{\infty} = g(\sigma^{\infty})$:
    \begin{align*}
        \sigma^{\infty}
        =\sigma^{\infty}\frac{\sigma_{\b{y}}}{\sigma^{\infty}+\sigma_{\b{y}}}+\nu,
    \end{align*}
    which is equivalent to $(\sigma^{\infty})^2-\sigma^{\infty}\nu -\nu\sigma_{\b{y}}= 0$. The positive root is
    \begin{align*}
        \sigma^{\infty}=\frac{\nu+ \sqrt{\nu^2+4\nu\sigma_{\b{y}}}}{2},
    \end{align*}
    which is the unique fixed point in $[\nu, \infty)$.
\end{proof}
\medskip

{\bf Simplification of belief update rules.} Given that the belief variances converge to fixed values, we can simplify our belief update rules~\eqref{eq:bayesian_update}-\eqref{eq:social_update} by substituting the final variance value directly. This approach is commonly used in Kalman filtering when computational resources are limited~\cite{SHI20111693}. Specifically, the final covariance matrix value is employed to compute constant gains, allowing the use of a time-invariant infinite-horizon filter in place of the truly optimal but time-varying solution. As a result, the update equations take the following simplified form:
\begin{subequations} \label{eqn:simplified}
    \begin{align}
    &\mu^{(i)+}_j[k]
    =\mu^{(i)}_j[k]
    +\frac{\sigma^{\infty}}{\sigma^{\infty}+\sigma_{\b{y}}}(\b{y}[k]-\mu^{(i)}_j[k]),\\
    &\alpha^{(i)+}_j[k]
    =\frac{\alpha^{(i)}_j[k] 
    \exp(-(\b{y}[k]-\mu^{(i)}_j[k])^2)
    }{\sum_{i=1}^M \alpha^{(i)}_j[k] \exp(-(\b{y}[k]-\mu^{(i)}_j[k])^2)},\\
    &\mu^{(i)}_j[k+1] = \mu^{(i)+}_j[k] + f_{\mu}(\mu^{(i)+}_j[k],\mu^{(i)+}_{-j}[k]),\\
    &\alpha^{(i)}_j[k+1] = \alpha^{(i)+}_j[k] \cdot f_{\alpha}(\alpha^{(i)+}_j[k],\alpha^{(i)+}_{-j}[k])
    \end{align}
\end{subequations}
for all \( i \in \mathsf{M} \).

\subsection{The Mean Dynamics}
In this subsection, we look at the mean dynamics. We consider a typical case, which has been widely applied in the consensus literature~\cite{degroot1974reaching, golub2010naive}. The means are updated according to the averaging process described in~\cite{bu2023discerning}. Specifically, the update rule is given by 
\[
    f_{\mu}(\mu^{(i)+}_j[k], \mu^{(i)+}_{-j}[k]) \!:=\! \delta_{\mu}\!\!\!\! \sum_{\ell : (\ell, j) \in \mathcal{E}} \!\!\!\! w_{\ell j} (\mu^{(i)+}_\ell[k] \!-\! \mu^{(i)+}_j[k]),
\]
where \( \delta_{\mu} > 0 \) is the mean belief mixing rate. 

 In this case, the simplified dynamics in~\eqref{eqn:simplified} can be further simplified to get
\begin{subequations}
    \begin{align}
    \mu^{(i)}_j[k+1]
    &=\mu^{(i)}_j[k]
    +\frac{\sigma^{\infty}}{\sigma^{\infty}+\sigma_{\b{y}}}(\b{y}[k]-\mu^{(i)}_j[k])
    \nonumber\\+&\frac{\delta_{\mu} \sigma_{\b{y}}}{\sigma^{\infty}+\sigma_{\b{y}}}\sum_{\ell:(\ell,j)\in\mathcal{E}} w_{\ell j}(\mu^{(i)}_\ell[k]
    -\mu^{(i)}_j[k])  ,\\
    \alpha^{(i)+}_j[k]
    &=\frac{\alpha^{(i)}_j[k] 
    \exp(-(\b{y}[k]-\mu^{(i)}_j[k])^2)
    }{\sum_{i=1}^M \alpha^{(i)}_j[k] \exp(-(\b{y}[k]-\mu^{(i)}_j[k])^2)},\\    
    \alpha^{(i)}_j[k+1]
    &=\alpha^{(i)+}_j[k]\cdot f_{\alpha}(\alpha^{(i)+}_j[k],\alpha^{(i)+}_{-j}[k])
    \end{align}
\end{subequations}
We can rewrite the mean dynamics in vector form as 
\begin{align}\label{eq:mu_dynamics}
    \mu^{(i)}[k\!+\!1]\!=\!\Sigma(I\!+\!\delta_{\mu} W\!-\!\delta_{\mu} D)\mu^{(i)}[k]\!+\!(I\!-\!\Sigma) \mathds{1}\b{y}[k],
\end{align}
where
\begin{align*}
    \mu^{(i)}[k]
    =&\begin{bmatrix}
        \mu^{(i)}_1[k],
        \dots,
        \mu^{(i)}_N[k]
    \end{bmatrix}^{\top}, 
    \Sigma:= \frac{\sigma_{\b{y}}}{\sigma^{\infty}+\sigma_{\b{y}}} I,\\
     D
    :=&
    \begin{bmatrix}
        \displaystyle \sum_{\ell:(\ell,1)\in\mathcal{E}} w_{\ell 1} & \cdots & 0 \\
        \vdots & \ddots & \vdots \\
        0 & \cdots & \displaystyle \sum_{\ell:(\ell,N)\in\mathcal{E}} w_{\ell N}
    \end{bmatrix},
\end{align*}
and $\mathds{1}$ is the vector of ones. Here, $\mathds{1}$ is of dimension $N \times 1$, and $I$ is of dimension $N \times N$.
\medskip

In what follows, we establish that the mean dynamics converge in probability.
\begin{theorem}[Convergence of means]
\label{thm:conv_mean}
    Assume that $(W-D)\mathds{1}=0$ and $\rho(\Sigma(I+\delta_{\mu} W-\delta_{\mu} D))<1$, where $\rho(\cdot)$ denotes the spectral radius. Then, 
    $$
    \mu^{(i)}[k] \overset{p}{\to} \mathcal{N}\left( \mathds{1}\theta,\; \frac{\sigma_{\b{y}}\sigma^{\infty}}{\sigma^{\infty}+2\sigma_{\b{y}}}\mathds{1}\mathds{1}^\top \right) \quad \text{as } k \to \infty.
    $$
\end{theorem}
\medskip

\begin{proof}
    The combination of Bayesian and social updates preserves the Gaussianity of $\mu^{(i)}[k]$. In what follows, we analyze convergence through mean and covariance dynamics. 
    
    {\bf Mean analysis.} The mean expectation evolves as 
    \begin{align}\label{eq:mean_mu_dynamics}
        &\mathbb{E}\{\mu^{(i)}[k+1]\} \notag\\
        = \ &\mathbb{E}\{\Sigma(I+\delta_{\mu} W-\delta_{\mu} D)\mu^{(i)}[k]+(I-\Sigma) \mathds{1}\b{y}[k]\}, \notag\\
        = \ &\Sigma(I+\delta_{\mu} W-\delta_{\mu} D)\mathbb{E}\{\mu^{(i)}[k]\}
         +(I-\Sigma) \mathds{1}\theta.
    \end{align}
    The condition $\rho(\Sigma(I+\delta_{\mu} W-\delta_{\mu} D))<1$ ensures that the system is stable and the mean dynamics converge to a steady state.  We solve the steady-state equation
    $$\overline{\mu}^{(i)}_{\infty} = \Sigma(I+\delta_{\mu} W-\delta_{\mu} D) \overline{\mu}^{(i)}_{\infty} + (I-\Sigma) \mathds{1}\theta,$$ which 
    yields  $\lim_{k\rightarrow\infty} \mathbb{E}\{\mu^{(i)}[k]\}=\mathds{1}\theta$. 
    
    {\bf Covariance analysis.} We define $$\tilde{\mu}^{(i)}[k] = \mu^{(i)}[k]-\mathbb{E}\{\mu^{(i)}[k]\}.$$ Combining Eqs.~\eqref{eq:mu_dynamics} and~\eqref{eq:mean_mu_dynamics}, we calculate that 
    \begin{align}
        & \tilde{\mu}^{(i)}[k+1] \notag \\
        = \ & \mu^{(i)}[k+1]-\mathbb{E}\{\mu^{(i)}[k+1]\} \notag \\
        = \ &  \Sigma(I+\delta_{\mu} W-\delta_{\mu} D)\tilde{\mu}^{(i)}[k] + (I-\Sigma) \mathds{1}(\b{y}[k] - \theta)
    \end{align}
    The covariance matrix is $P[k]=\mathbb{E}\{\tilde{\mu}^{(i)}[k]\tilde{\mu}^{(i)}[k]^\top\}$, which evolves as
    \begin{align}\label{eq:covariance_dynamics}
        P[k+1]
        = \ & \mathbb{E}\{\tilde{\mu}^{(i)}[k+1]\tilde{\mu}^{(i)}[k+1]^\top\} \notag \\
        = \ &\Sigma(I+\delta_{\mu} W-\delta_{\mu} D) P[k] (I+\delta_{\mu} W-\delta_{\mu} D)^\top \Sigma^\top  \notag
        \\&+ \sigma_{\b{y}}(I-\Sigma )\mathds{1}\mathds{1}^\top (I-\Sigma ).
    \end{align}
    The condition $\rho(\Sigma(I+\delta_{\mu} W-\delta_{\mu} D))<1$ ensures that the system is stable and the covariance dynamics converge to a steady state. We solve the steady-state equation
    \begin{align*}
        P_{\infty} 
        =&\Sigma(I+\delta_{\mu} W-\delta_{\mu} D) P_{\infty} (I+\delta_{\mu} W-\delta_{\mu} D)^\top \Sigma^\top 
        \\&+ \sigma_{\b{y}}(I-\Sigma )\mathds{1}\mathds{1}^\top (I-\Sigma ),
    \end{align*}
    which yields $\lim_{k\rightarrow\infty} P[k]=\frac{\sigma_{\b{y}}\sigma^{\infty}}{\sigma^{\infty}+2\sigma_{\b{y}}}\mathds{1}\mathds{1}^\top.$
   
   Therefore, the belief means $\mu[k]$ converge in probability to:
    $$
    \mu^{(i)}[k] \overset{p}{\to} \mathcal{N}\left(  \mathds{1}\theta,\; \frac{\sigma_{\b{y}}\sigma^{\infty}}{\sigma^{\infty}+2\sigma_{\b{y}}}\mathds{1}\mathds{1}^\top \right) \quad \text{as } k \to \infty.
    $$ This completes the proof.
\end{proof}
\medskip

Theorem~\ref{thm:conv_mean} shows that, if $(W-D)\mathds{1}=0$ and $\rho(\Sigma(I+\delta_{\mu} W-\delta_{\mu} D))<1$, the wisdom of crowd prevails, that is, the agents recover the ground truth as time grows. Particularly, because $\lim_{\nu\rightarrow 0}\lim_{k\rightarrow\infty} P[k]=0$, all the opinions concentrate on the ground truth $\theta$ if $\nu\ll 1$. 

Note that, by definition, $(W-D)\mathds{1}=0$ for all $W$ and $D$ matrices defined above if there are no self-loops present in the graph, i.e., $(j,j)\notin \mathcal{E}$ for all $j\in\mathcal{V}$. Therefore, $(W-D)\mathds{1}=0$ irrespective of the fact that the weights are positive or negative (i.e., agents are friends or foes). The signs and the weights, of course, can change the spectral radius of $I-\delta_{\mu} D+\delta_{\mu} W$ and impact our ability to meet the condition that $\rho(\Sigma(I-\delta_{\mu} D+\delta_{\mu} W))<1$. 

\medskip

\begin{remark}
    Sharp readers may notice that multi-modal beliefs collapse into a single-modal belief when all modes share the same mean expectation. This happens because the observation signal follows a simple Gaussian likelihood rather than a mixture, and the Bayesian update rule treats all modes equally regardless of the observation signal's distance from their means. However, if we modify the update rule so that closer ones contribute more while distant ones contribute less, the mode means may remain distinct instead of merging into the same. We leave the exploration of this modified update rule to future research.
\end{remark}
\medskip

\begin{remark}
    An important next step is to investigate different social update policies for weights, which could provide deeper insights into the mechanisms governing opinion shifts.
    For instance, we can assume that the social update policy for weights is given by
\begin{align*}
    f_{\alpha}(\alpha^{(i)+}_j[k],\alpha^{(i)+}_{-j}[k])
    &\propto \prod_{\ell:(\ell,j)\in\mathcal{E}}\left(\frac{\alpha^{(i)+}_\ell[k]}{\alpha^{(i)+}_j[k]}\right)^{w_{\ell j}}.
\end{align*}
Based on Theorem~\ref{thm:conv_mean}, $\mu^{(i)}[k]$ would converge in probability to $\mathcal{N}\left( \mathds{1}\theta,\; \frac{\sigma_{\b{y}}\sigma^{\infty}}{\sigma^{\infty}+2\sigma_{\b{y}}}\mathds{1}\mathds{1}^\top \right)$. We can approximate the weight update to
\begin{align*}
    \alpha^{(i)}_j[k+1]
    \propto &\alpha^{(i)}_j[k]\exp(-(\b{y}[k]-\mu^{(i)}_j)^2) 
    \\
    &\times \prod_{\ell:(\ell,j)\in\mathcal{E}}\left( \frac{\alpha^{(i)}_j[k] 
    }{ \alpha^{(i)}_j[k] 
    } \right)^{w_{\ell j}}.
\end{align*}
Taking logarithms from both sides of the equation, we get a linear update rule for the logarithm of the weights. This equation can be investigated to understand the weight equilibrium. We also leave this exploration to future research.
\end{remark}
\section{Persuasion and Centrality}\label{sec:stubborn}
In this section, we focus on the presence of a stubborn agent in the network \cite{bu2023discerning}. Assume that the first agent\footnote{This choice is without loss of generality up to renumbering the agents} is a stubborn agent that does not change its belief, i.e., $\mu^{(i)}_{1}[k]=\mu^{\dagger}$ for all $k\in\mathbb{N}$. 

Let $b \in \mathbb{R}^{N}$ be a vector. We denote $b_{-1} \in \mathbb{R}^{N-1}$ as the vector obtained by excluding its first entry. Let $A \in \mathbb{R}^{N \times N}$ be a matrix. We employ a block matrix representation:
\[
A = \begin{bmatrix} A_{1,1} & A_{1,-1} \\ A_{-1,1} & A_{-1,-1} \end{bmatrix},
\]
where $A_{1,-1}$ represents the submatrix obtained by removing the first column from the first row of $A$, $A_{-1,1}$ represents the submatrix obtained by removing the first row while retaining only the first column, and $A_{-1,-1}$ corresponds to the submatrix obtained by excluding both the first row and the first column of $A$. 

Now define $A := \Sigma(I+\delta_{\mu} W-\delta_{\mu} D)$ and $B:= (I-\Sigma)$. The mean dynamics in the presence of a stubborn agent can be rewritten as
\begin{align*}
    \begin{bmatrix}
        \mu^{(i)}_1[k+1]
        \\
        \mu^{(i)}_{-1}[k+1]
    \end{bmatrix}=&
    \begin{bmatrix}
    0 & 0\\
        A_{-1,1} & A_{-1,-1}
    \end{bmatrix} \begin{bmatrix}
        \mu^{\dagger}
        \\
        \mu^{(i)}_{-1}[k]
    \end{bmatrix} \\
    &+     \begin{bmatrix}
    1 & 0\\
        0 & B_{-1,-1}
    \end{bmatrix} \begin{bmatrix}
        \mu^{\dagger}\\
        \mathds{1}_{N-1}\b{y}[k]
    \end{bmatrix}.
\end{align*}
The mean of malleable agents can be rewritten as
\begin{align*}
    \mu^{(i)}_{-1}[k+1]
    &\!=\!A_{-1,1}\mu^{\dagger}\!+\!A_{-1,-1}\mu^{(i)}_{-1}[k] \!+\! B_{-1,-1}\mathds{1}_{N-1}\b{y}[k].
\end{align*}

\begin{theorem}[Convergence with stubborn agents]\label{thm:stubborn_mean} Assume that  $\rho(A_{-1,-1})<1$. Then, $\lim_{k\rightarrow\infty} \mathbb{E}\{\mu^{(i)}_{-1}[k]\}= \gamma$,
where 
\begin{equation}\label{eq:stubborn_mean}
    \gamma \!=\! (I_{N-1}\!-\!A_{-1,-1})^{-1}\left(B_{-1,-1}\mathds{1}_{N-1}\theta\!+\!A_{-1,1}\mu^{\dagger}\right).
\end{equation}
\end{theorem}
\medskip
\begin{proof}
    In what follows, we conduct mean analysis. The mean expectation evolves as  
    \begin{align*}
     &\mathbb{E}\{\mu^{(i)}_{-1}[k+1]\}\\
    =\ &A_{-1,1}\mu^{\dagger}+A_{-1,-1}\mathbb{E}\{\mu^{(i)}_{-1}[k]\} + B_{-1,-1}\mathds{1}_{N-1}\theta.   
    \end{align*}
    The condition $\rho(A_{-1,-1})<1$ ensures that the system is stable and the dynamics converge. We solve the following steady-state equation
    $$\overline{\mu}^{(i)}_{-1\infty}
        = A_{-1,1}\mu^{\dagger}+A_{-1,-1}\overline{\mu}^{(i)}_{-1\infty} + B_{-1,-1}\mathds{1}_{N-1}\theta,   $$
    which yields $\lim_{k\rightarrow\infty} \mathbb{E}\{\mu^{(i)}_{-1}[k]\}=\gamma.$
\end{proof}
\medskip

The value of $\lim_{k\rightarrow\infty} \mathbb{E}\{\mu^{(i)}_{-1}[k]\}$ in Theorem~\ref{thm:stubborn_mean} shows the influence that the stubborn agent has on discovering media bias. It therefore captures the ease with which the stubborn agent can manipulate the information flow and can be used as a measure of centrality in the social network.


\begin{table}[b]
    \centering
    \begin{tabular}{ccccc}
    	
        \hline
        \multicolumn{5}{c}{\textbf{Four Settings}} \\
        \hline
    	& S1 & S2 & S3 & S4 \\
    	
    	Stubborn? & False & False & True & True \\
        
        $\mu^{\dagger}$ & N/A & N/A & -1 & -1\\
    	
    	$\theta$ & 1 & - & - & - \\
    	
    	$\sigma_{\mathbf{y}}$ & 0.1 & 1 & 0.1 & 1 \\
    	
    	$\delta_{\mu}$ & 0.6 & - & - & - \\
    	
    	$\delta_{\sigma}$ & 0.1 & - & - & - \\
    	
    	$\nu$ & 0.1 & - & - & - \\
    	
    	$\mu^{(1)}_{j}[0]$ & $\mathcal{U}_{[0,1]}$ & - & - & - \\
    	
    	$\mu^{(2)}_{j}[0]$ & $\mathcal{U}_{[-1,0]}$ & - & - & - \\
    	
    	$\sigma^{(1)}_{j}[0]$ & 1 & - & - & - \\
    	
    	$\sigma^{(2)}_{j}[0]$ & 1 & - & - & - \\
        \hline
    \end{tabular}
    \caption{The initialization for four settings S1-S4, where $\mathcal{U}_{[a,b]}$ represents uniform distribution over the interval $[a,b]$ and ``-'' represents the same as that of S1.}
    \label{tab:parameter_settings}
\end{table}
\section{Numerical Simulation}\label{sec:simulation}
In this section, we verify the convergence results and investigate the influence of the stubborn agent. The code is open-sourced at \url{https://github.com/chyj528/Model-Multi-Modal-Beliefs-}.

{\bf Network setup.} We consider a network of 50 nodes based on the Watts–Strogatz small-world model. First, we generate a Watts–Strogatz network with an average node degree of $k_{ws} = 3$ and a rewiring probability of $p_{ws} = 0.2$. To introduce a relatively \textit{influential} node, we modify the network by connecting node 1 to half of the remaining nodes.  Once the network is realized, each existing link $(i, j) \in \mathcal{E}$ is assigned a weight $w_{ij}$, generated randomly from a uniform distribution between 0 and 1. In the simulation, while the adjacency matrix is symmetric, the weight matrix is not, meaning $w_{ij} \neq w_{ji}$.

{\bf Model setup.} In the considered network, the news media organization has a true bias represented by the parameter $\theta$, which is used as the basis for releasing articles to the public. Each node models the media organization's true bias using a two-mode Gaussian mixture $\bs{\theta} \sim \mathcal{M}(\{\mu^{(i)}_j[0]\}_{i=1}^2, \{\sigma^{(i)}_j[0]\}_{i=1}^2, \{\alpha^{(i)}_j[0]\}_{i=1}^2)$. Each node observes the news media's true bias through articles according to a Gaussian distribution $\b{y}[k] \sim \mathcal{N}(\bs{\theta}, \sigma_{\b{y}})$. Each node estimates the ground truth bias $\bs{\theta}$ based on its observations of the sample articles and the belief exchanges within the  network. Each node first updates its belief according to its observations of the sample articles, following  the Bayesian update rule Eq.~\eqref{eq:bayesian_update}, and then exchanges its belief over network, following the social update rule Eq.~\eqref{eq:social_update}. The social update policies are characterized by 
\begin{subequations}
\begin{align}
    f_{\mu}&= \delta_{\mu} \sum_{\ell : (\ell, j) \in \mathcal{E}} w_{\ell j} (\mu^{(i)+}_\ell[k] - \mu^{(i)+}_j[k]),\\
    f_{\sigma}&= \delta_{\sigma} \sum_{\ell : (\ell, j) \in \mathcal{E}} w_{\ell j} (\sigma^{(i)+}_\ell[k] - \sigma^{(i)+}_j[k]) + \nu. \label{eq:f_sigma}
\end{align}
\end{subequations}

{\bf Initialization.} We consider four settings: one without stubborn agents, where the observation variance is relatively small; one without stubborn nodes, where the observation variance is relatively large; one with stubborn nodes, where the observation variance is small; and one with stubborn nodes, where the observation variance is large. The parameter values and initial belief values are shown in Table~\ref{tab:parameter_settings}.
\medskip

\begin{figure}[tb]
    \centering
    \includegraphics[width=0.925\linewidth]{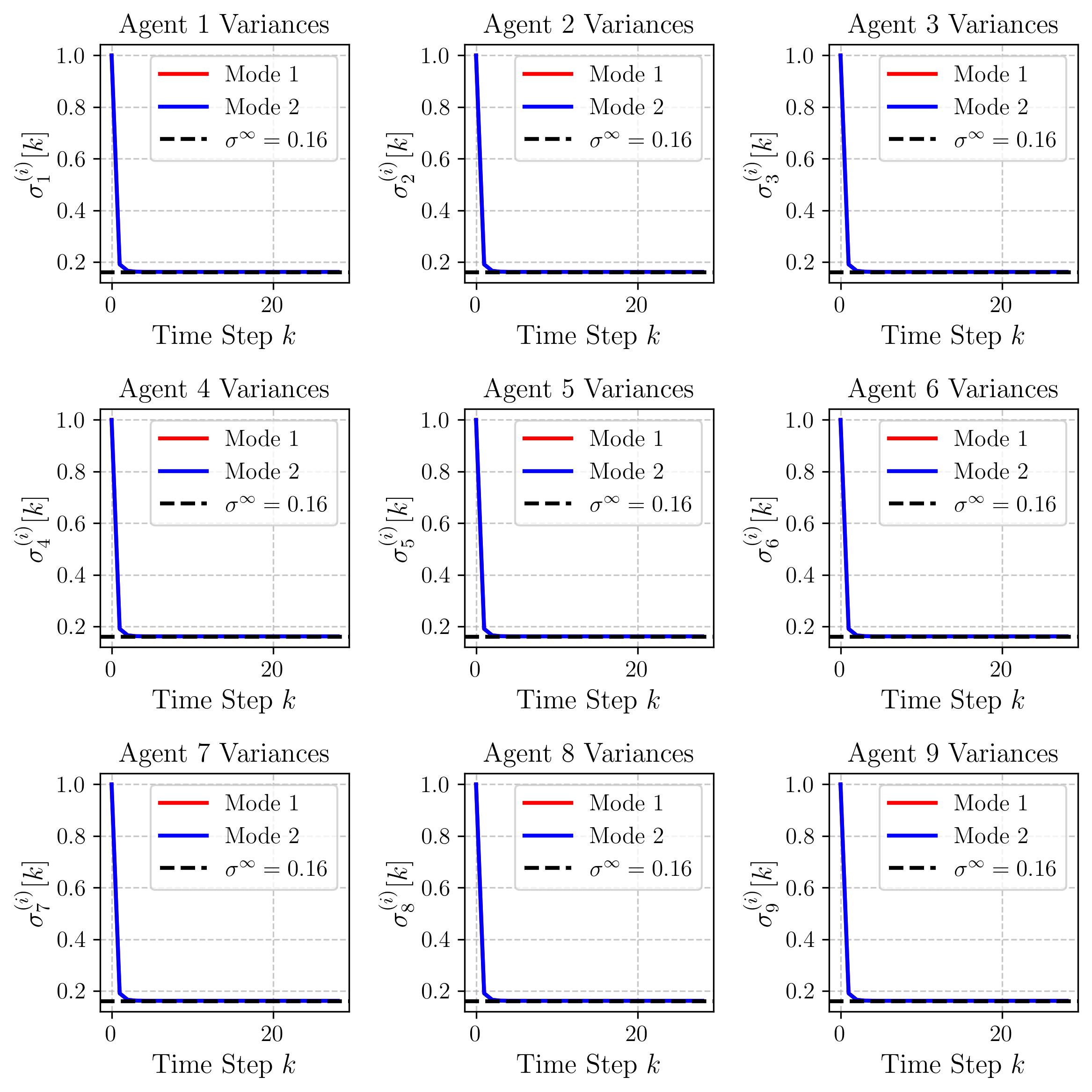}
    \caption{The variance trajectories of the first nine nodes under setting S1.}
    \label{fig:variance_traj}
\end{figure}

\begin{figure}[tb]
    \centering
    \includegraphics[width=0.925\linewidth]{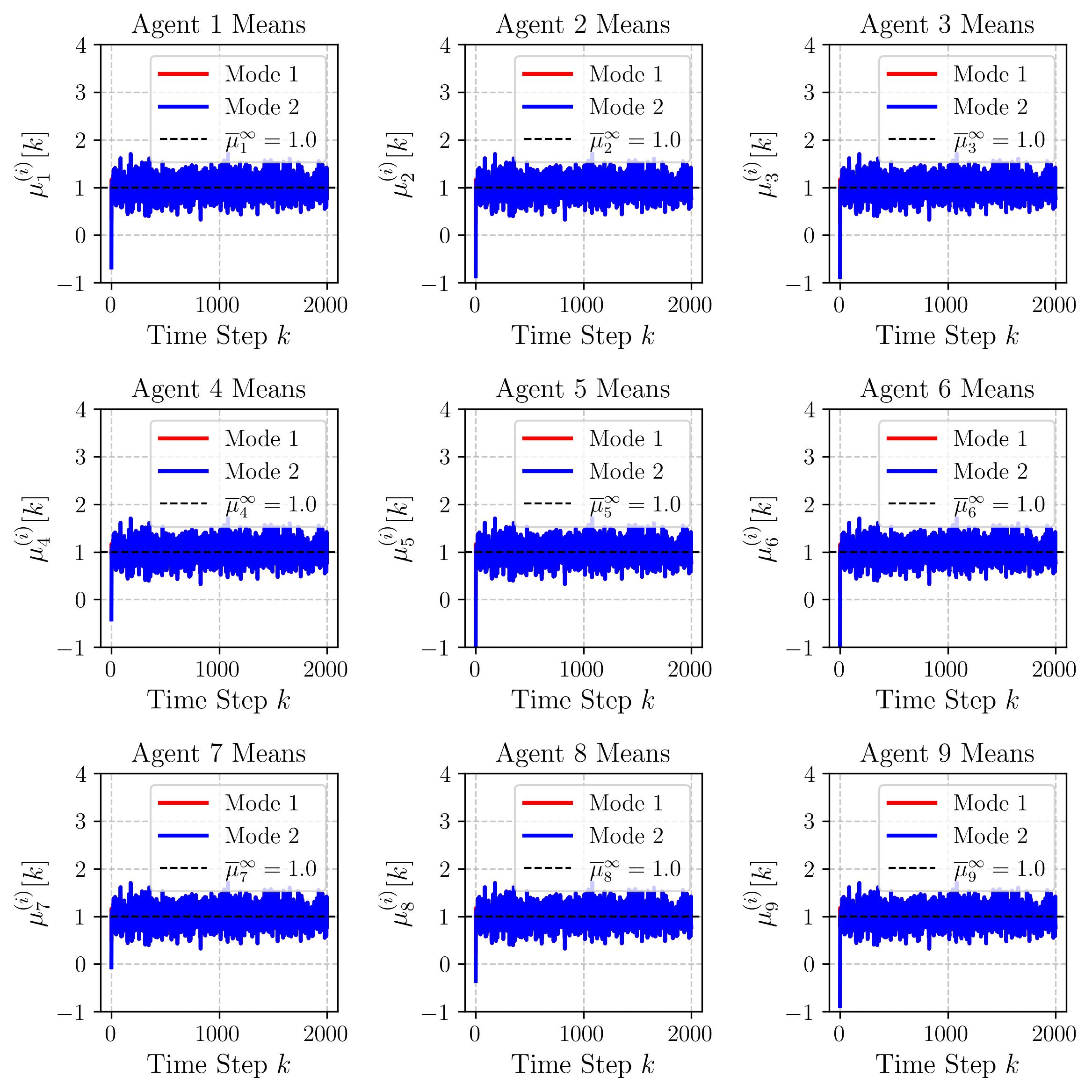}
    \caption{The mean trajectories of the first nine nodes under setting S1.}
    \label{fig:ns_mean_traj}
\end{figure}

\begin{figure}[tb]
    \centering
    \includegraphics[width=0.925\linewidth]{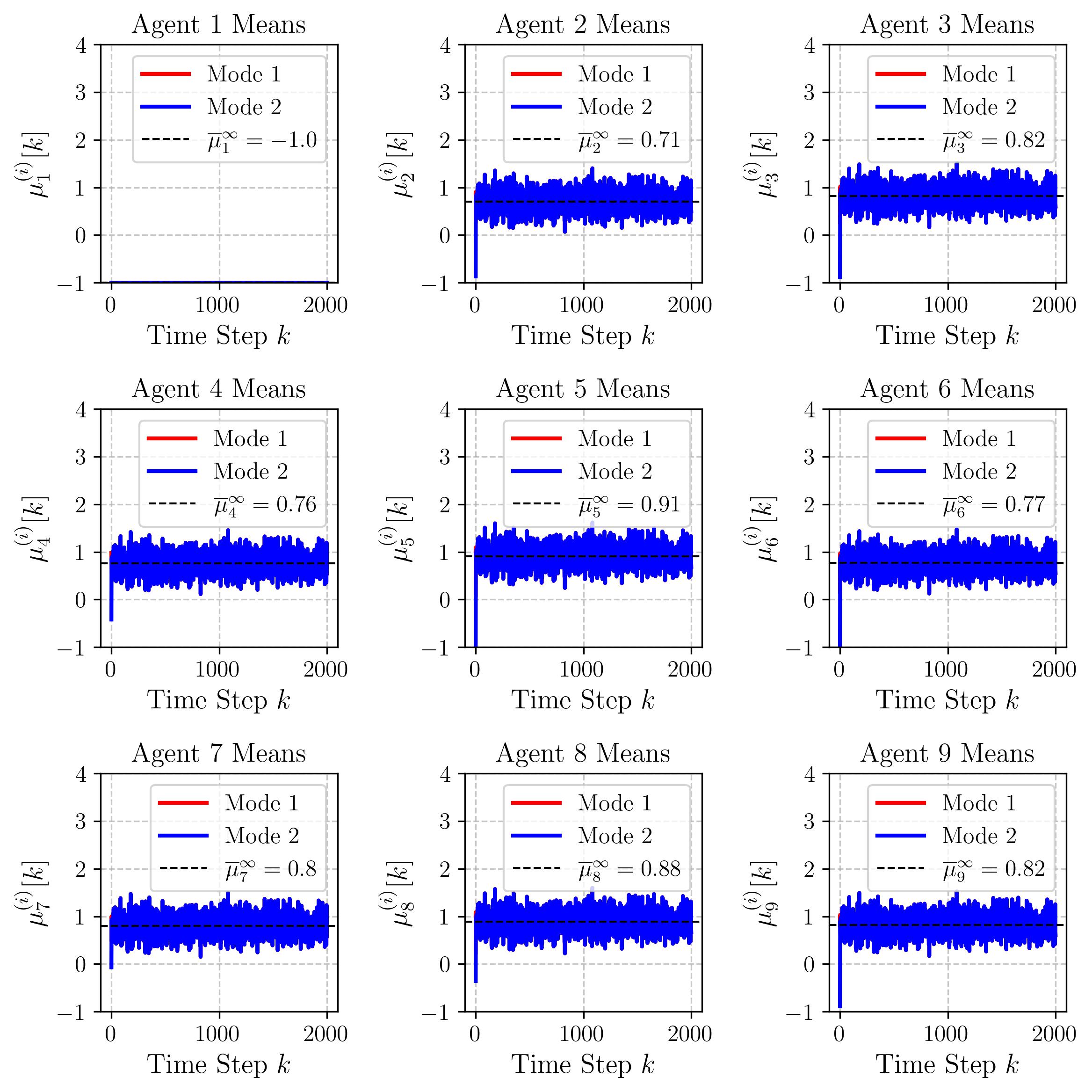}
    \caption{The mean trajectories of the first nine nodes under setting S3.}
    \label{fig:s_mean_traj}
\end{figure}

\begin{experiment}[Validation of Theorems~\ref{thm:BAA}-\ref{thm:conv_mean}]
In this experiment, we consider there is no stubborn agent under setting S1. We first look at the variance dynamics. From Eq.~\eqref{eq:f_sigma}, we observe that $f_{\sigma} = \nu$ when all variables take identical values, satisfying the assumptions of Theorem~\ref{thm:BAA}. Additionally, S1 in Table~\ref{tab:parameter_settings} shows that the initial variances are identical, satisfying the assumptions of Theorem~\ref{thm:BAA}. Fig.~\ref{fig:variance_traj} presents the variance trajectories of the first nine nodes, all of which converge to the theoretical value $\sigma^{\infty}$, computed from Eq.~\eqref{eq:sigma_infty}, thus validating Theorem~\ref{thm:BAA}. 

It is noted that the rapid convergence of variance dynamics allows for the direct substitution of final variance values, simplifying the update equations  (see Eq.~\eqref{eqn:simplified}). Next, we examine the mean dynamics. We verify that $(W - D) \mathds{1} = 0$ and that $\rho(\Sigma(I + \delta_{\mu} W - \delta_{\mu} D)) < 1$, confirming that the assumptions in Theorem~\ref{thm:conv_mean} hold. Fig.~\ref{fig:ns_mean_traj} illustrates the mean trajectories of the first nine nodes, showing that the expected mean converges to the inherent bias of the news media organization, thus validating Theorem~\ref{thm:conv_mean}.
\end{experiment}
\medskip

\begin{experiment}[Validation of Theorem~\ref{thm:stubborn_mean}]
    In this experiment, we consider there is a stubborn agent under setting S3. The stubborn agent maintains its belief at $\mu^{\dagger} = -1$. We adopt the same process in Experiment 1. We verify that $\rho(A_{-1,-1}) < 1$, satisfying the assumption in Theorem~\ref{thm:stubborn_mean}.  Fig.~\ref{fig:s_mean_traj} shows the mean trajectories of the first nine nodes, showing that the expected mean converges to the theoretical values calculated from Eq.~\eqref{eq:stubborn_mean}, thus validating Theorem~\ref{thm:stubborn_mean}.
\end{experiment}
\medskip
\begin{figure}[tb]
\centering
\subfloat[Setting S2]{\includegraphics[width=0.65\linewidth]{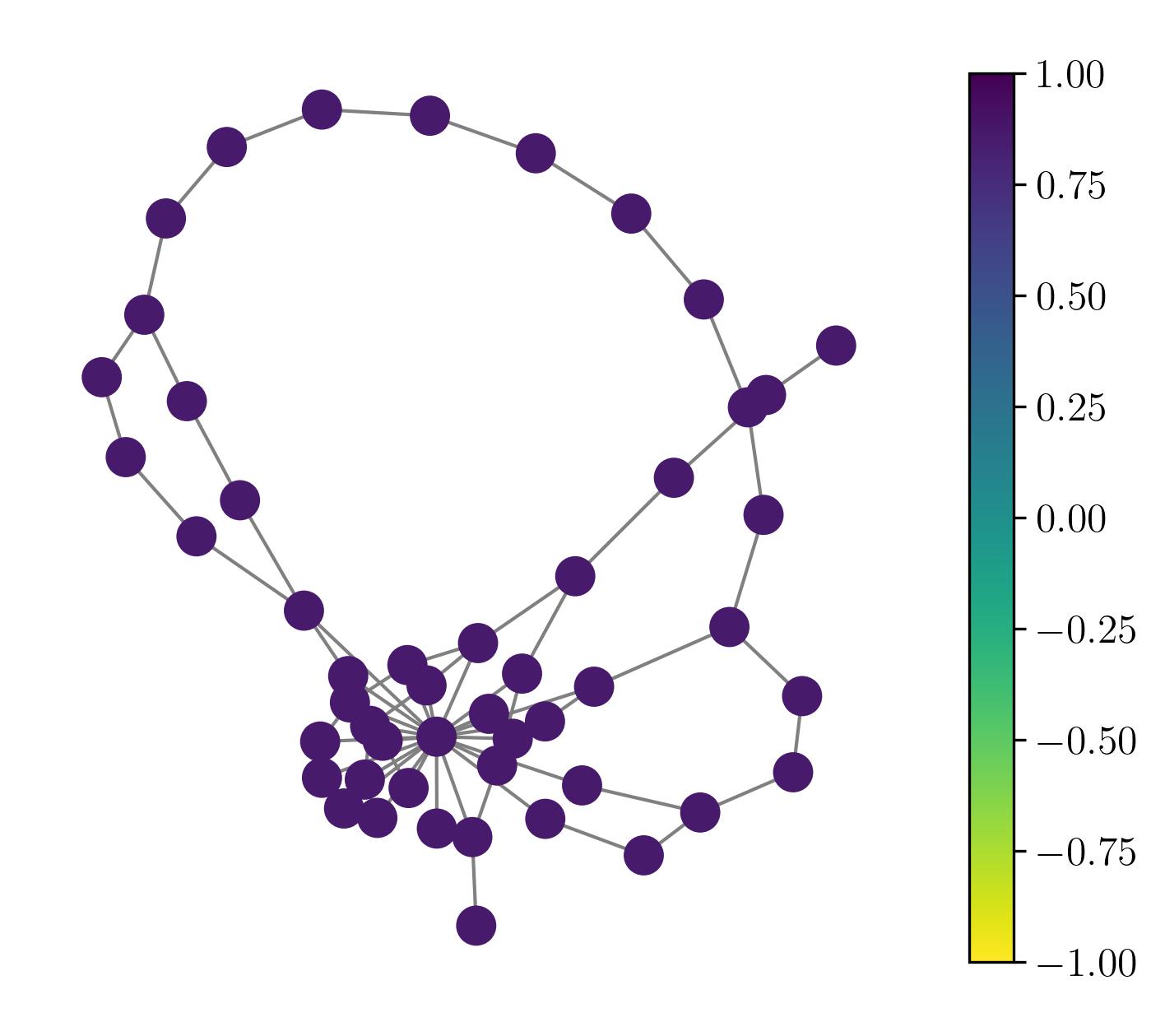}}\\
\subfloat[Setting S3]{\includegraphics[width=0.65\linewidth]{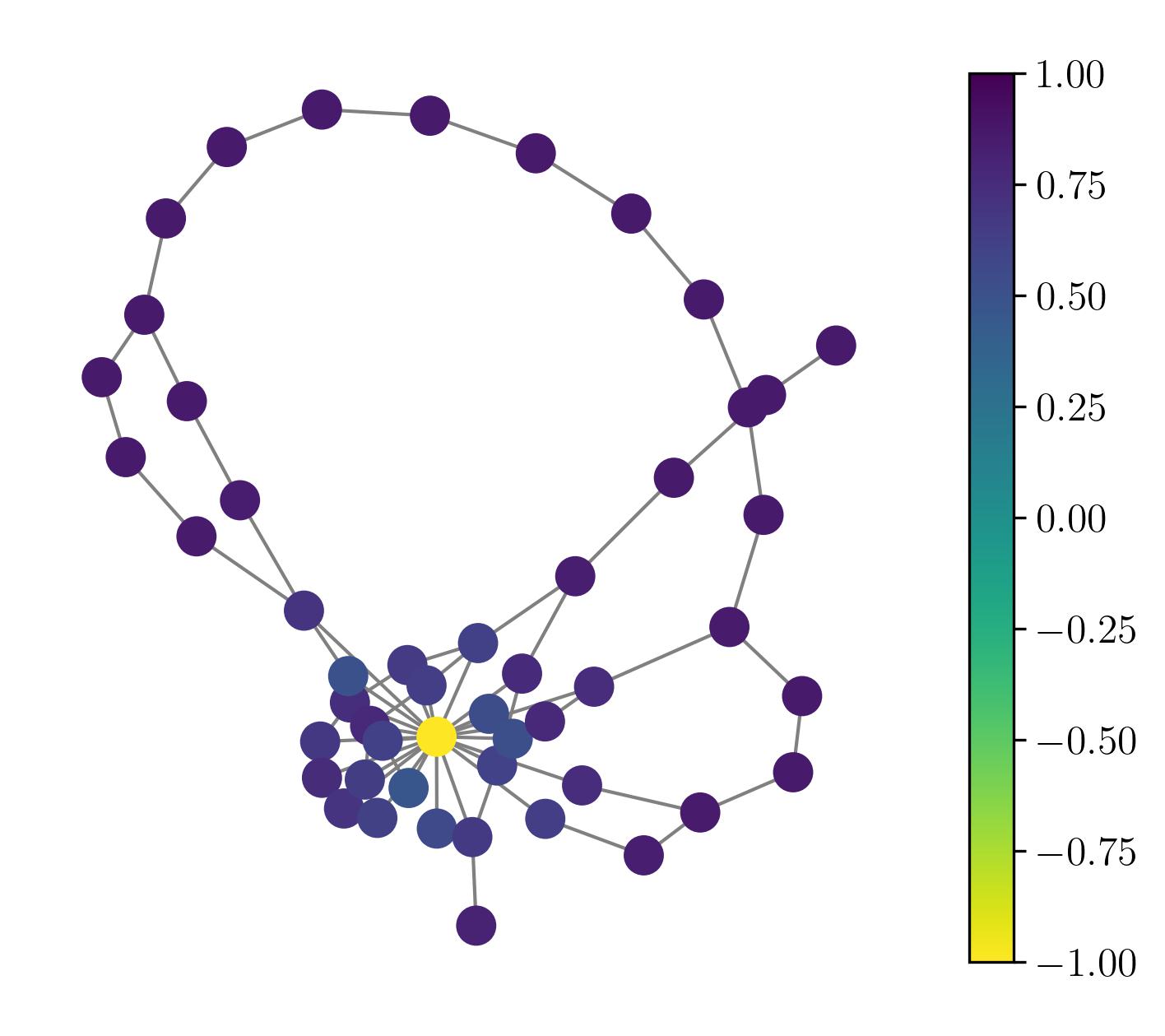}}\\
\subfloat[Setting S4]{\includegraphics[width=0.65\linewidth]{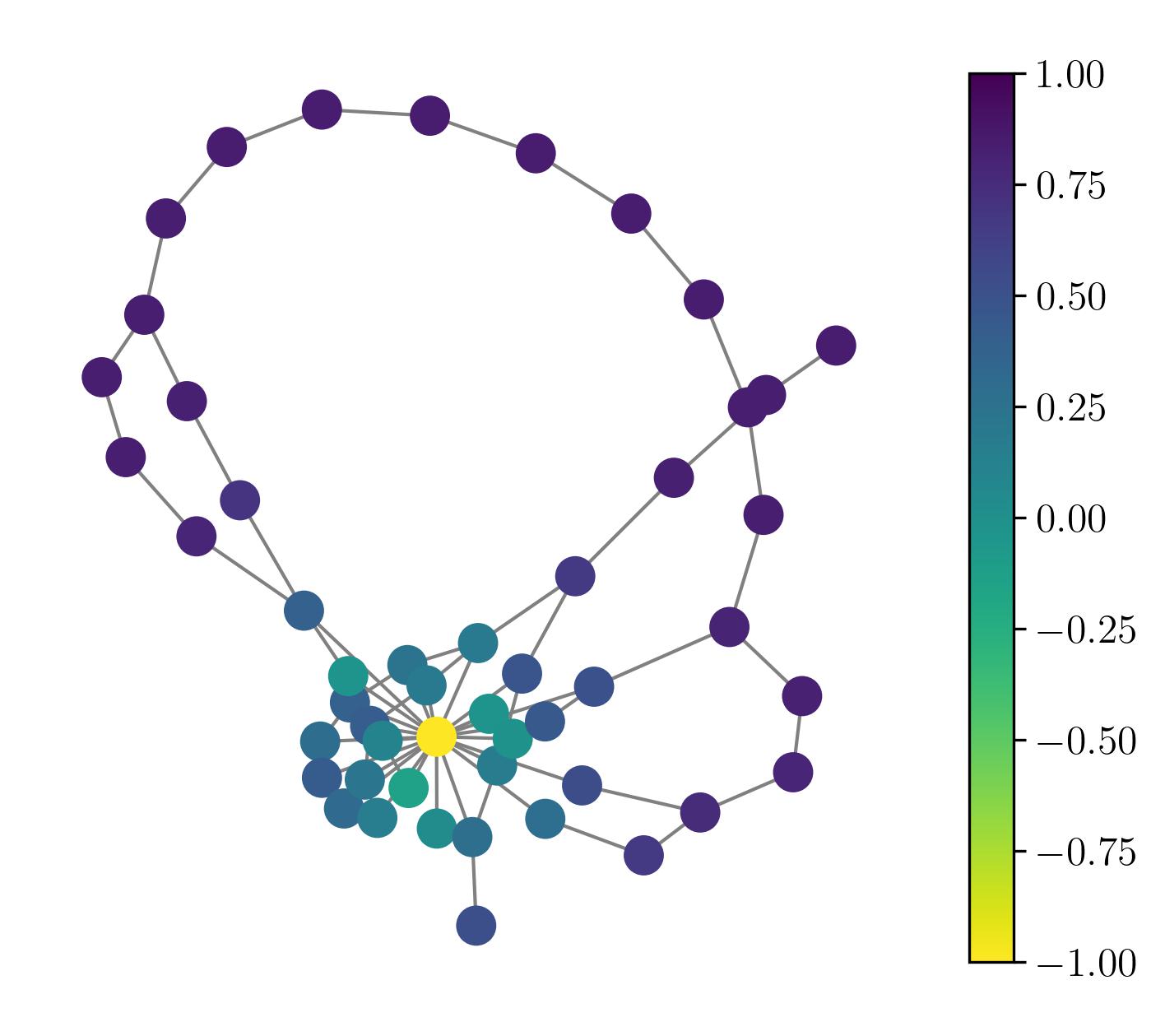}}
\caption{The mean belief at the equilibrium $\lim_{k\rightarrow\infty} \mathbb{E}\{\mu^{(i)}[k]\}$ under setting S2, S3, and S4.}
\label{fig:graph_cmap}
\end{figure}

\begin{experiment}[Investigation of Stubborn Node's Effect]
In this experiment, we consider three settings S2, S3, and S4 to investigate the effect of a stubborn node with $\mu^{\dagger} = -1$ on the ability to discern the true bias $\theta = -1$ of the news media organization under different observation variances. We run the belief update process for 2000 time steps and compute the mean average over the last 1000 time steps to approximate the mean expectation equilibrium, $\lim_{k\rightarrow\infty} \mathbb{E}\{\mu^{(i)}[k]\}$.  We use colormaps to plot graphs for the three settings in Fig.~\ref{fig:graph_cmap}, where the color of each node represents its corresponding mean expectation equilibrium. The stubborn agent, which maintains a fixed belief over time, is highlighted in yellow in Figs.~\ref{fig:graph_cmap}(b)–\ref{fig:graph_cmap}(c). 

Comparing Fig.~\ref{fig:graph_cmap}(a) with Figs.~\ref{fig:graph_cmap}(b)–\ref{fig:graph_cmap}(c), we observe that the presence of the stubborn node negatively impacts the ability of other nodes to correctly discern media bias. Furthermore, comparing Fig.~\ref{fig:graph_cmap}(b) with Fig.~\ref{fig:graph_cmap}(c), we find that when the observed media signal $\b{y}[k]$ has a small variance, i.e., $\frac{\sigma_{\b{y}}}{\sigma^{\infty}} \ll 1$, the nodes can identify the bias with reasonable accuracy. However, when the media signal is highly uncertain, i.e., $\frac{\sigma_{\b{y}}}{\sigma^{\infty}} \gg 1$, the stubborn node exerts greater influence, preventing some nodes from correctly discerning the true media bias.

\end{experiment}

\section{Conclusions and Future Work}
\label{sec:conc}
In this work, we employed Gaussian mixtures to model multi-modal beliefs and opinion uncertainty in the context of opinion dynamics across social networks. Through preliminary theoretical results and numerical simulations, we explored how opinion formation and uncertainty are influenced by stubborn individuals acting as social influencers. Future research can extend this work in several directions. One promising avenue is to generalize observation signals by adopting a Gaussian mixture likelihood instead of a simple Gaussian likelihood. Additionally, an important next step is to investigate different multiplicative dynamics for the weight evolution within the Gaussian mixture model, which could provide deeper insights into the mechanisms governing opinion shifts.


\bibliographystyle{IEEEtran}
\bibliography{ref}

\begin{thebibliography}{10}
\providecommand{\url}[1]{#1}
\csname url@samestyle\endcsname
\providecommand{\newblock}{\relax}
\providecommand{\bibinfo}[2]{#2}
\providecommand{\BIBentrySTDinterwordspacing}{\spaceskip=0pt\relax}
\providecommand{\BIBentryALTinterwordstretchfactor}{4}
\providecommand{\BIBentryALTinterwordspacing}{\spaceskip=\fontdimen2\font plus
\BIBentryALTinterwordstretchfactor\fontdimen3\font minus
  \fontdimen4\font\relax}
\providecommand{\BIBforeignlanguage}[2]{{%
\expandafter\ifx\csname l@#1\endcsname\relax
\typeout{** WARNING: IEEEtran.bst: No hyphenation pattern has been}%
\typeout{** loaded for the language `#1'. Using the pattern for}%
\typeout{** the default language instead.}%
\else
\language=\csname l@#1\endcsname
\fi
#2}}
\providecommand{\BIBdecl}{\relax}
\BIBdecl

\bibitem{druckman2005impact}
J.~N. Druckman and M.~Parkin, ``The impact of media bias: How editorial slant
  affects voters,'' \emph{Journal of Politics}, vol.~67, no.~4, pp. 1030--1049,
  2005.

\bibitem{gonzalez2023social}
S.~Gonz{\'a}lez-Bail{\'o}n and Y.~Lelkes, ``Do social media undermine social
  cohesion? a critical review,'' \emph{Social Issues and Policy Review},
  vol.~17, no.~1, pp. 155--180, 2023.

\bibitem{viswanath2007mass}
K.~Viswanath, S.~Ramanadhan, and E.~Z. Kontos, ``Mass media,'' in
  \emph{Macrosocial determinants of population health}.\hskip 1em plus 0.5em
  minus 0.4em\relax NY: Springer, 2007, pp. 275--294.

\bibitem{french1956formal}
J.~R.~P. French, ``A formal theory of social power,'' \emph{Psychological
  Review}, vol.~63, no.~3, p. 181, 1956.

\bibitem{degroot1974reaching}
M.~H. DeGroot, ``Reaching a consensus,'' \emph{Journal of the American
  Statistical Association}, vol.~69, no. 345, pp. 118--121, 1974.

\bibitem{friedkin1997social}
N.~E. Friedkin and E.~C. Johnsen, ``Social positions in influence networks,''
  \emph{Social networks}, vol.~19, no.~3, pp. 209--222, 1997.

\bibitem{carletti2006make}
T.~Carletti, D.~Fanelli, S.~Grolli, and A.~Guarino, ``How to make an efficient
  propaganda,'' \emph{Europhysics Letters}, vol.~74, no.~2, p. 222, 2006.

\bibitem{pineda2015mass}
M.~Pineda and G.~M. Buend{\'\i}a, ``Mass media and heterogeneous bounds of
  confidence in continuous opinion dynamics,'' \emph{Physica A}, vol. 420, pp.
  73--84, 2015.

\bibitem{amelkin2017polar}
V.~Amelkin, F.~Bullo, and A.~K. Singh, ``Polar opinion dynamics in social
  networks,'' \emph{IEEE Transactions on Automatic Control}, vol.~62, no.~11,
  pp. 5650--5665, 2017.

\bibitem{baumann2020modeling}
F.~Baumann, P.~Lorenz-Spreen, I.~M. Sokolov, and M.~Starnini, ``Modeling echo
  chambers and polarization dynamics in social networks,'' \emph{Physical
  Review Letters}, vol. 124, no.~4, p. 048301, 2020.

\bibitem{9992374}
L.~Wang, G.~Chen, Y.~Hong, G.~Shi, and C.~Altafini, ``A social power game for
  the concatenated {Friedkin-Johnsen} model,'' in \emph{Proceedings of {IEEE}
  Conference on Decision and Control}, 2022, pp. 3513--3518.

\bibitem{jia2015opinion}
P.~Jia, A.~MirTabatabaei, N.~E. Friedkin, and F.~Bullo, ``Opinion dynamics and
  the evolution of social power in influence networks,'' \emph{SIAM Review},
  vol.~57, no.~3, pp. 367--397, 2015.

\bibitem{low2022discerning}
N.~K.~Y. Low and A.~Melatos, ``Discerning media bias within a network of
  political allies and opponents: The idealized example of a biased coin,''
  \emph{Physica A}, vol. 590, p. 126722, 2022.

\bibitem{anunrojwong2018naive}
J.~Anunrojwong and N.~Sothanaphan, ``Naive bayesian learning in social
  networks,'' in \emph{Proceedings of {ACM} 19th Conference on Economics and
  Computation}, 2018, pp. 619--636.

\bibitem{jadbabaie2012non}
A.~Jadbabaie, P.~Molavi, A.~Sandroni, and A.~Tahbaz-Salehi, ``Non-bayesian
  social learning,'' \emph{Games and Economic Behavior}, vol.~76, no.~1, pp.
  210--225, 2012.

\bibitem{bu2023discerning}
Y.~Bu and A.~Melatos, ``Discerning media bias within a network of political
  allies and opponents: Disruption by partisans,'' \emph{Physica A}, vol. 624,
  p. 128958, 2023.

\bibitem{golub2010naive}
B.~Golub and M.~O. Jackson, ``Naive learning in social networks and the wisdom
  of crowds,'' \emph{American Economic Journal: Microeconomics}, vol.~2, no.~1,
  pp. 112--149, 2010.

\bibitem{flam2012mmse}
J.~T. Flam, S.~Chatterjee, K.~Kansanen, and T.~Ekman, ``On {MMSE} estimation: A
  linear model under {Gaussian} mixture statistics,'' \emph{IEEE Transactions
  on Signal Processing}, vol.~60, no.~7, pp. 3840--3845, 2012.

\bibitem{4518754}
A.~Kundu, S.~Chatterjee, A.~Sreenivasa~Murthy, and T.~V. Sreenivas, ``{GMM}
  based {Bayesian} approach to speech enhancement in signal/transform domain,''
  in \emph{2008 IEEE International Conference on Acoustics, Speech and Signal
  Processing}, 2008, pp. 4893--4896.

\bibitem{low2022vacillating}
N.~K.~Y. Low and A.~Melatos, ``Vacillating about media bias: Changing one’s
  mind intermittently within a network of political allies and opponents,''
  \emph{Physica A}, vol. 604, p. 127829, 2022.

\bibitem{SHI20111693}
L.~Shi, P.~Cheng, and J.~Chen, ``Sensor data scheduling for optimal state
  estimation with communication energy constraint,'' \emph{Automatica},
  vol.~47, no.~8, pp. 1693--1698, 2011.

\end{thebibliography}
	
\end{document}